\newtheorem{theorem}{Theorem}

\newtheorem{definition}[theorem]{Definition}

\documentclass[12pt]{article}

\begin{document}

\title{Conformal Actions in Any Dimension}
\author{Andr\'{e} Wehner\thanks{%
Department of Physics, Utah State University, Logan, Utah 84322} \thanks{%
Electronic mail: sllg7@cc.usu.edu} and James T. Wheeler\thanks{%
Department of Physics, Utah State University, Logan, Utah 84322} \thanks{%
Electronic mail: jwheeler@cc.usu.edu}}
\maketitle

\begin{abstract}
Biconformal gauging of the conformal group gives a scale-invariant volume
form, permitting a single form of the action to be invariant in any
dimension. We display several $2n$-dim scale-invariant polynomial actions
and a dual action. We solve the field equations for the most general action
linear in the curvatures for a minimal torsion geometry. In any dimension $%
n>2$, the solution is foliated by equivalent $n$-dim Ricci-flat Riemannian
spacetimes, and the full $2n$-dim space is symplectic. Two fields defined
entirely on the Riemannian submanifolds completely determine the solution: a
metric $e_{\mu }^{a},$ and a symmetric tensor $k_{\mu \nu }$.
\end{abstract}

\section{Introduction}

One of the problems facing the use of the conformal group as a fundamental
spacetime symmetry in $n$\ dimensions is the highly restricted set of
possible actions. In sharp contrast to general relativity, where the
Einstein-Hilbert action is Lorentz and coordinate invariant in every
dimension, conformal actions are typically coupled to the dimension. This
coupling to dimension occurs because under a rescaling of the metric by a
factor $e^{2\phi }$, the volume element of an $n$-dimensional spacetime
scales by $e^{n\phi }$.\ Therefore, for example, since an action containing $%
k$\ copies of the scale-invariant conformal tensor, $C_{bcd}^{a}$, requires $%
k$\ inverse metrics (each scaling as $e^{-2\phi }$) to form a Lorentz
scalar, we find that an expression such as
\[
S=\int \sqrt{-g}C_{b\mu \nu }^{a}C_{c\alpha \beta }^{b}\cdots C_{a\rho
\sigma }^{c}\ g^{\mu \alpha }g^{\nu \rho }\cdots g^{\beta \sigma }\ d^{n}x
\]
is scale invariant only if $n-2k=0$.

A number of techniques have been developed to overcome this problem, most of
them within the context of conformal gauge theory. Generally, these
techniques either require additional ``compensating'' fields or fail to
reproduce general relativity in any gauge. Here we show that because of its
scale-invariant volume form, the biconformal gauging of the conformal group
\cite{New Conformal Gauging Paper} resolves these problems, allowing us to
write an invariant action linear in the curvature without compensating
fields. We begin our discussion with a brief overview of some of the
previous treatments of conformal gauging.

The gauging of the conformal group in four dimensions has been handled in
much the same way as Poincar\'{e} gauging, simply treating the dilations and
special conformal transformations as generators of additional symmetries. As
recounted in \cite{Romao+Ferber+Freund}, it was long believed that special
conformal transformations were ``ungaugeable'' because the conformal matter
current is explicitely $x$-dependent, so that coupling it to the special
conformal gauge field would spoil translation invariance. Therefore, prior
to 1977, conformal gauging incorporated only Lorentz transformations,
dilations and, for the superconformal group SU(2,2|N), the
internal U(N) symmetry algebras, i.e., Weyl's theory of gravity was regarded
as the unique gauge theory of the conformal group. In order to remain as
close as possible to Einstein's theory, Deser \cite{Deser} coupled a
massless Lorentz scalar field $\phi (x)$ (dilaton) of compensating conformal
weight $-1$ to gravitation through the manifestly scale-invariant quantity $%
\frac{1}{6}\phi ^{2}R$. Later, Dirac \cite{Dirac}, trying to
accommodate the Large Numbers Hypothesis, similarly modified Weyl's free
Lagrangian by replacing all $R^{2}$-type terms by $\phi ^{2}R$. This method
gave rise to various theories involving the ``generalized'' Einstein
equations \cite{Freund}-\cite{Bicknell}. They were shown to reduce to
general relativity when expressed in a particular gauge
\cite{Freund}-\cite{Bramson}.

In 1977, it was demonstrated by Crispim-Romao, Ferber and Freund \cite
{Romao+Ferber+Freund} and independently by Kaku, Townsend and Van
Nieuwenhuizen \cite{Kaku+Townsend}-\cite{Kaku+Townsend2} that special
conformal transformations can indeed be gauged. Using a Weyl-like
conformally invariant $4$-dimensional action quadratic in the conformal
curvatures and the assumption of vanishing torsion, it is found that the
gauge fields associated with special conformal transformations are
algebraically removable. The action reduces to a scale-invariant,
torsion-free Weyl theory of gravity based on the square of the conformal
curvature. This auxiliary nature of the special conformal gauge field has
been shown to follow for any $4$-dimensional action quadratic in the
curvatures \cite{Auxiliary Field}. Generically, the action reduces to the a
linear combination of the square of the conformal curvature and the square
of the curl of the Weyl vector.

Alternatively, the special conformal gauge fields may be removed by the
curvature constraint \cite{Townsend+van Nieuwenhuizen}
\begin{equation}
R_{bac}^{a}=0  \label{Constraint 1}
\end{equation}
This ensures that $\mathbf{R}_{b}^{a}$ is just the Weyl conformal curvature
tensor, rather than the usual Riemann curvature. Then the constraint of
vanishing torsion,
\begin{equation}
\mathbf{T}^{a}=0  \label{Constraint 2}
\end{equation}
also renders the spin connection auxiliary. We will refer to conditions (\ref
{Constraint 1}) and (\ref{Constraint 2}) as the conventional constraints.
The dilation field (Weyl vector) drops out of the action completely, so
instead of a Weyl unified theory one again obtains a $4$-dimensional
Weyl-trivial theory of gravity, gauge equivalent to a Riemannian geometry.

The quadratic curvature theory was later generalized to dimensions $n>4$ by
unifying it with the compensating field approach \cite{Bergshoef}. The
action involves terms of the form
\[
e_{[a}^{\ \mu }e_{b}^{\ \nu }e_{c}^{\ \alpha }e_{d]}^{\ \beta }\phi
^{2(n-4)/(n-2)}R_{\mu \nu }^{\quad ab}R_{\alpha \beta }^{\quad cd}
\]
While the resulting field equations no longer require the special conformal
gauge fields to be removable, the conventional constraints may still be
imposed to remove them. These constraints were shown to be conformally
invariant if the conformal weight of $\phi $ is $-(n-2)/2$. Unfortunately,
this class of theories is not equivalent to general relativity in any gauge.
It is useful, though, in the understanding of superconformal gauge theories
in $n=6$ and $10$\ dimensions \cite{BergSalam+Sezgin}.

A different use of a compensating field proves somewhat more successful. In $%
n$ dimensions, an action of the form $\phi \fbox{} \phi $ is conformally
invariant. Because the conformal d'Alembertian contains a term involving the
trace of the special conformal gauge field, constraint (\ref{Constraint 1})
leads to a $\phi ^{2}R$ term in the action. Again imposing the conventional
constraints, and gauging the Weyl vector to zero and the compensating field
to a constant, we recover Einstein gravity in $n$ dimensions.

For $n=3,$ the Chern-Simons form leads to an exactly soluble (super-)
conformal gravity theory \cite{Deser+Jackiw+Templeton}-\cite{Witten}
characterized by conformal flatness.

In $n=2$ dimensions, the conformal group is not a Lie group; it is generated
by the infinite-dimensional Virasoro algebra \cite
{Belavin+Polyakov+Zamolodchikov}. The importance of $2$-dimensional
conformal field theory is well known as the symmetry of the $2$-dimensional
world sheet in string theory \cite{Friedan}. In addition, we note the recent
surge of interest in conformal field theories due to the celebrated AdS/CFT
duality conjecture put forward by Maldacena \cite{Maldacena} and made more
precise by others \cite{Gubser}, which relates type IIB string theory and M
theory in certain $(n+1)$-dimensional anti-de-Sitter spacetime backgrounds
to conformally invariant field theories in $n$ dimensions.

Recently, a new way of gauging the conformal group \cite{New Conformal
Gauging Paper} has been proposed which resolves the problem of writing
scale-invariant actions in arbitrary dimension \textit{without} using
compensating fields. In particular, we write the most general \textit{linear}
action and find that all minimal torsion solutions are foliated by
equivalent $n$-dimensional Ricci-flat Riemannian spacetimes. Thus, the new
gauging establishes a clear connection between conformal gauge theory and
general relativity. It does \textit{not} require the conventional
constraints.

The new gauging starts with the conformal group\ that acts on an $n$-dim
spacetime. We will always assume $n>2$\ and can thus identify the conformal
group with the $\frac{1}{2}(n+1)(n+2)$-parameter orthogonal group $O(n,2)$,
which acts on an $n$-dimensional compact spacetime and leaves the null
interval $ds^{2}=\eta _{\mu \nu }dx^{\mu }dx^{\nu }=0$, $\eta _{\mu \nu
}=diag(1\ldots 1,-1,-1)$, $\mu ,\nu =1\ldots n$, invariant. It is generated
by Lorentz transformations, dilations (rescalings), translations, and
special conformal transformations\footnote{{\small These transformations
have elsewhere been called proper conformal transformations or conformal
boosts. See Appendix A for a formal definition of the conformal group and an
overview of geometrical gauge theory.}}. The latter are actually
translations of the inverse coordinate $y_{\mu }\equiv -\eta _{\mu \nu }%
\frac{x^{\nu }}{x^{2}}$, or, equivalently, translations of the vertex of the
lightcone at infinity. In the new gauging, they are treated on an equal
footing with translations and in that context will be referred to as\textit{%
\ co-translations}. We retain the term special conformal transformations for
conformal gauge theories in which one of the two subsets of $n$
transformations (i.e., translations or special conformal transformations) is
treated differently from the other.

By demanding that the translational and co-translational gauge fields
together span the base manifold the biconformal technique yields a $2n$%
-dimensional manifold. A summary of this technique is given in Appendix A
and full detail is available in reference \cite{New Conformal Gauging Paper}%
. Among the advantages to this procedure is the fact that the resulting
volume element is scale-invariant. To see why, notice that the inverse
coordinates $y_{\mu }$\ scale oppositely to the spacetime coordinates $%
x^{\mu }.$\ The corresponding translational and co-translational gauge
fields, $\mathbf{\omega }^{a}$\ and $\mathbf{\ \omega }_{a}$\ scale as $%
e^{\phi }$\ and $e^{-\phi },$\ respectively. The volume element,
\[
\mathbf{\omega }^{a_{1}}\wedge \ldots \wedge \mathbf{\omega }^{a_{n}}\wedge
\mathbf{\omega }_{b_{1}}\wedge \ldots \wedge \mathbf{\omega }_{b_{n}},
\]
is therefore scale invariant, since there are $n$\ translations and $n$\
co-transla-tions. The scale invariance of the volume form eliminates the
typical coupling of invariance to dimension, opening up a wide range of
possible actions.

In the next section, we define our notational conventions.

In Sec.(3) we introduce the zero-weight biconformal Levi-Civita tensor,
define the biconformal dual, present a large class of polynomial actions for
biconformal geometries valid for all dimensions $n>2$, and write the most
general action linear in the biconformal curvatures and structural
invariants. Finally, we note certain topological integrals.

The subsequent three sections examine the consequences of the most general
linear action found in Sec.(3). We completely solve the field equations for
a minimal torsion biconformal space, and show that the solutions admit two
foliations of the $2n$-dim base manifold. The first involution shows that
the base space is foliated by conformally flat $n$-dim submanifolds. The
second involution gives a foliation by equivalent $n$-dim spacetimes
constrained by the vanishing of the Ricci tensor. Thus, the solder form
satisfies the vacuum Einstein equation despite the overall geometry
possessing (minimal) torsion, a non-trivial Weyl vector, and an arbitrary
cosmological constant. Each Riemannian geometry is fully determined by the
components of the solder form, $e_{\mu }^{\;a},$\ defined entirely on these
spacetime submanifolds. The full $2n$-dim solution contains one additional
field, a symmetric tensor $h_{\mu \nu },$ also defined entirely on the
submanifolds. Except for a single special case, the full $2n$-dim space is
necessarily symplectic, hence almost complex and almost K\"{a}hler.

Sec.(7) treats one special case which occurs in the general solution. In
this case, the Ricci tensor continues to vanish while certain additional
fields are allowed. Sec.(8) compares and contrasts the present method with
previous conformal and scale-invariant gaugings, while the final section
contains a brief summary.

\section{Notation}

The group $O(n,2)$\ preserves the $(n+2)$-dim metric $diag(1\ldots
1,-1,-1)$, or equivalently in a null basis
\[
\eta _{\tilde{A}\tilde{B}}=\left(
\begin{array}{lll}
\;\;0 & 0\cdots 0 & \;\;1 \\
\begin{array}{l}
0 \\
\ \vdots \\
0
\end{array}
& \eta _{ab} &
\begin{array}{l}
0 \\
\ \vdots \\
0
\end{array}
\\
\;\;1 & 0\cdots 0 & \;\;0
\end{array}
\right)
\]
where $\tilde{A},\tilde{B},\ldots =0,1,\ldots ,n,n+1$\ and $a,b,\ldots
=1,\ldots ,n$. The Minkowski metric is written as $\eta _{ab}=$\ $%
diag(1\ldots 1,-1)$. The usual antisymmetry of the pseudo-orthonormal
connection, $\mathbf{\omega }_{\tilde{B}}^{\tilde{A}},$ allows us to express
$\mathbf{\omega }_{A}^{n+1},$\ $\mathbf{\omega }_{n+1}^{A},$\ and $\mathbf{\
\omega }_{n+1}^{n+1}$\ (where $A,B,\ldots =0,1,\ldots ,n$) in terms of the
remaining set,
\[
\mathbf{\omega }_{B}^{A}=\{\mathbf{\omega }_{b}^{a},\mathbf{\omega }%
^{a}\equiv \mathbf{\omega }_{0}^{a},\mathbf{\omega }_{a}\equiv \mathbf{\
\omega }_{a}^{0},\mathbf{\omega }_{0}^{0}\}
\]
These remaining independent connection components (gauge fields) are
associated with the Lorentz, translation, co-translation, and dilation
generators of the conformal group, respectively. We refer to $\mathbf{\omega
}_{b}^{a}$\ as the spin-connection, $\mathbf{\omega }^{a}$\ as the
solder-form, $\mathbf{\omega }_{a}$\ as the co-solder-form, and$\mathbf{\
\omega }_{0}^{0}$\ as the Weyl vector, where in all cases differential forms
are bold and the wedge product is assumed between adjacent forms. The $%
O(n,2) $\ curvature, given by
\begin{equation}
\mathbf{\Omega }_{B}^{A}=\mathbf{d\omega }_{B}^{A}-\mathbf{\omega }_{A}^{C}%
\mathbf{\omega }_{C}^{A},  \label{O(n,2) curvature}
\end{equation}
divides into corresponding parts, $\{\mathbf{\Omega }_{b}^{a},\mathbf{\Omega
}^{a}\equiv \mathbf{\Omega }_{0}^{a},\mathbf{\Omega }_{a}\equiv \mathbf{\
\Omega }_{a}^{0},\mathbf{\Omega }_{0}^{0}\}$, called the\ curvature,\
torsion, co-torsion, and\ dilation, respectively. While these parts are not
conformally invariant, they are invariant under the fiber symmetry of the
biconformal bundle.

When broken into parts based on the homogeneous Weyl transformation
properties of the biconformal bundle, i.e. Lorentz and scale
transformations, eq.(\ref{O(n,2) curvature}) becomes
\begin{eqnarray}
\mathbf{\Omega }_{b}^{a} &=&\mathbf{d\omega }_{b}^{a}-\mathbf{\omega }%
_{b}^{c}\mathbf{\omega }_{c}^{a}-\Delta _{cb}^{ad}\mathbf{\omega }_{d}%
\mathbf{\omega }^{c}  \label{Structure 1} \\
\mathbf{\Omega }^{a} &=&\mathbf{d\omega }^{a}-\mathbf{\omega }^{b}\mathbf{%
\omega }_{b}^{a}-\mathbf{\omega }_{0}^{0}\mathbf{\omega }^{a}
\label{Structure 2} \\
\mathbf{\Omega }_{a} &=&\mathbf{d\omega }_{a}-\mathbf{\omega }_{a}^{b}%
\mathbf{\omega }_{b}-\mathbf{\omega }_{a}\mathbf{\omega }_{0}^{0}
\label{Structure 3} \\
\mathbf{\Omega }_{0}^{0} &=&\mathbf{d\omega }_{0}^{0}-\mathbf{\omega }^{a}%
\mathbf{\omega }_{a}  \label{Structure 4}
\end{eqnarray}
where $\Delta _{cd}^{ab}\equiv \delta _{c}^{a}\delta _{d}^{b}-\eta ^{ab}\eta
_{cd}$. If we set $\mathbf{\omega }_{a}=0=\mathbf{\Omega }_{a}$\ the
equations reduce to the structure equations of an $n$-dim Weyl geometry with
torsion. Each curvature may be expanded in the $(\mathbf{\omega }^{a},%
\mathbf{\omega }_{b})$\ basis as
\[
\mathbf{\Omega }_{B}^{A}=\frac{1}{2}\Omega _{Bcd}^{A}\mathbf{\omega }%
^{cd}+\Omega _{Bd}^{Ac}\mathbf{\omega }_{c}\mathbf{\omega }^{d}+\frac{1}{2}%
\Omega _{B}^{Acd}\mathbf{\omega }_{cd}
\]
where we introduce the convention of writing $\mathbf{\omega }^{bc\ldots
d}\equiv \mathbf{\omega }^{b}\mathbf{\omega }^{c}\ldots \mathbf{\omega }^{d}=%
\mathbf{\omega }^{b}\wedge \mathbf{\omega }^{c}\wedge \ldots \wedge \mathbf{%
\omega }^{d}$\ etc. The three terms will be called the spacetime-, cross-,
and momentum-term, respectively, of the corresponding curvature. For each
curvature of the set $\{\mathbf{\Omega }_{b}^{a},\mathbf{\Omega }^{a},%
\mathbf{\Omega }_{a}\}$, each of these three terms is a distinct
Weyl-covariant object. Each term of the dilation $\mathbf{\Omega }_{0}^{0}$\
is Weyl invariant. In addition, the $2$-forms $\mathbf{d\omega }_{0}^{0}\ $
and $\mathbf{\omega }^{a}\mathbf{\omega }_{a}$\ appearing in eq.(\ref
{Structure 4}) are Weyl invariant.

In working with biconformal objects it is simpler to abandon the raising and
lowering of indices with the metric, for two reasons. First, it would lead
to confusion of fields that are independent, such as the spacetime and
cross-terms of the curvature, $\Omega _{bcd}^{a}$\ and$\mathbf{\ }\Omega
_{bd}^{ac}$, or the necessarily independent $1$-forms, $\mathbf{\omega }^{a}$
\ and $\mathbf{\omega }_{a}.$\ Second, the position of any lower-case Latin
index corresponds to the associated scaling weight: each upper index
contributes $+1$ to the weight, while each lower index contributes $-1.$
Thus, $\Omega _{bcd}^{a}$ \ has weight $-2,$\ while $\Omega _{a}^{\;bc}$\
has weight $+1.$

\section{Biconformal Actions}

In order to construct biconformal actions, we must first examine certain
special properties of the volume element of a biconformal space. Since the
base manifold is spanned by the $2n$\ gauge fields $\{\mathbf{\omega }^{a},%
\mathbf{\omega }_{b}\}$\ we may set
\[
\mathbf{\Phi }=\varepsilon _{a_{1}\cdots a_{n}}^{\qquad b_{1}\cdots b_{n}}%
\mathbf{\omega }^{a_{1}}\cdots \mathbf{\omega }^{a_{n}}\mathbf{\omega }
_{b_{1}}\cdots \mathbf{\omega }_{b_{n}}
\]
where $\varepsilon _{a_{1}\cdots a_{n}}^{\qquad \quad b_{1}\cdots b_{n}}$\
is the $2n$-dim Levi-Civita symbol. The mixed index positioning indicates
the scaling weight of the indices, and \underline{not}\ any use of the
metric. The positions arise from our notation for $O(n,2)$, in which the
generators have a pair of indices, $L_{B}^{A}.$

The Lorentz transformations of the biconformal fields are $n$- rather than $%
2n$-dim matrices\footnote{{\small While the matrix structure of a Lorentz
transformation in biconformal space is }$n${\small -dim, the\ functional
dependence is }$2n${\small -dim. Thus, we have }$\Lambda _{b}^{a}=\Lambda
_{b}^{a}(x^{\mu },y_{\nu })${\small , where }$(x^{\mu },y_{\nu })${\small \
span the full biconformal space.}}. Therefore, the two $n$-dim Levi-Civita
symbols are also Lorentz invariant, but have opposite scaling properties,
\begin{eqnarray*}
\varepsilon ^{a_{1}\cdots a_{n}} &\rightarrow &e^{n\phi }\varepsilon
^{a_{1}\cdots a_{n}} \\
\varepsilon _{a_{1}\cdots a_{n}} &\rightarrow &e^{-n\phi }\varepsilon
_{a_{1}\cdots a_{n}}
\end{eqnarray*}
Another way to see the intrinsic presence of these two tensors is from the
distinguishability of the $\mathbf{\omega }^{a}$\ from the $\mathbf{\omega }%
_{a}$ by their differing scale weights, giving rise to two $n$-volumes,
\begin{eqnarray*}
\mathbf{\Phi }_{+} &=&\varepsilon _{a_{1}\cdots a_{n}}\mathbf{\omega }%
^{a_{1}}\cdots \mathbf{\omega }^{a_{n}} \\
\mathbf{\Phi }_{-} &=&\varepsilon ^{a_{1}\cdots a_{n}}\mathbf{\omega }%
_{a_{1}}\cdots \mathbf{\omega }_{a_{n}}
\end{eqnarray*}
Therefore,
\[
\mathbf{\Phi }=\mathbf{\Phi }_{+}\mathbf{\Phi }_{-}
\]
and we may set
\[
\varepsilon _{a_{1}\cdots a_{n}}^{\qquad b_{1}\cdots b_{n}}=\varepsilon
_{a_{1}\cdots a_{n}}\varepsilon ^{b_{1}\cdots b_{n}}
\]
which is clearly both Lorentz and scale invariant. Whenever the indices on $%
\varepsilon _{a_{1}\cdots a_{n}}^{\qquad b_{1}\cdots b_{n}}$\ are not in
this standard position, the signs are to be adjusted accordingly. Thus, for
example,
\[
\varepsilon _{a\;c\cdots d\qquad b}^{\;g\qquad e\cdots f\;h}\equiv
-\,\varepsilon _{abc\cdots d}^{\qquad e\cdots fgh}
\]
The Levi-Civita tensor is normalized such that traces are given by
\[
\varepsilon _{a_{1}\cdots a_{p}c_{p+1}\cdots c_{n}}\varepsilon ^{b_{1}\cdots
b_{p}c_{p+1}\cdots c_{n}}=p!(n-p)!\delta _{a_{1}\cdots a_{p}}^{b_{1}\cdots
b_{p}}
\]
where the totally antisymmetric $\delta $-symbol is defined as
\[
\delta _{a_{1}\cdots a_{p}}^{b_{1}\cdots b_{p}}\equiv \delta
_{a_{1}}^{[b_{1}}\cdots \delta _{a_{p}}^{b_{p}]}
\]

The Levi-Civita tensor allows us to define biconformal duals of forms. Let
\[
T\equiv \frac{1}{p!q!}T_{a_{1}\cdots a_{p}}^{\qquad b_{1}\cdots b_{q}}%
\mathbf{\omega }^{a_{1}\cdots a_{p}}\mathbf{\omega }_{b_{1}\cdots b_{q}}
\]
be an arbitrary $(p+q)$-form, $p,q\in \{0,\ldots ,n\},$\ with weight $p-q$.
Then the dual of $T$\ is a $((n-q)+(n-p))$-form, also of weight $%
(n-q)-(n-p)=p-q,$\ defined as
\[
^{\ast }T\equiv \frac{1}{p!q!(n-p)!(n-q)!}T_{a_{1}\cdots a_{p}}^{\qquad
b_{1}\cdots b_{q}}\varepsilon _{b_{1}\cdots b_{n}}^{\qquad \ a_{1}\cdots
a_{n}}\mathbf{\omega }^{b_{q+1}\cdots b_{n}}\mathbf{\omega }_{a_{p+1}\cdots
a_{n}}
\]
so that
\[
^{\ast *}T=(-1)^{(p+q)(n-(p+q))}T\
\]
Note that the dual map is scale-invariant: both $T$\ and $^{*}T$\ have
weight $(p-q)$.

We can now write a variety of biconformally invariant and $O(n,2)$\
invariant integrals.

To build biconformal invariants, we use the fact that the fiber symmetry
(structure group) of the biconformal bundle is the Weyl group, consisting of
Lorentz transformations and dilations, while the connection forms
corresponding to translations and co-translations span the base space. We
can therefore return to the reduced notation, and find a correspondingly
increased number of possible actions. First, we note that the bilinear form
\[
\mathbf{\omega }^{a}\mathbf{\omega }_{b}
\]
is scale invariant. This object allows us to write actions of arbitrary
order, $k=1,\ldots ,n,$\ in the curvatures. In particular, we can write
\[
S_{m,k-m}=\int \mathbf{\Omega }_{B_{1}}^{A_{1}}\ldots \mathbf{\Omega }
_{B_{m}}^{A_{m}}\mathbf{\Omega }_{0}^{0}\ldots \mathbf{\Omega }_{0}^{0}%
\mathbf{\omega }^{a_{1}\cdots a_{n-k}}\mathbf{\omega }_{b_{1}\cdots
b_{n-k}}Q_{A_{1}\cdots A_{m}a_{1}\cdots a_{n-k}}^{B_{1}\cdots
B_{m}b_{1}\cdots b_{n-k}}
\]
where there are $m$\ factors of the curvature $\Omega _{B}^{A}$\ and $k-m$\
factors of the dilational curvature, $\Omega _{0}^{0}$. The invariant tensor
$Q_{A\cdots C}^{B\cdots D}$\ has $2[n-(k-m)]$ \ indices and must be built
from $\delta _{B}^{A}$\ and the Levi-Civita tensor. Notice that when $m=k$\
we can use
\[
Q_{A_{1}\cdots A_{k}a_{1}\cdots a_{n-k}}^{B_{1}\cdots B_{k}b_{1}\cdots
b_{n-k}}=\varepsilon ^{0B_{1}\cdots B_{k}b_{1}\cdots b_{n-k}}\varepsilon
_{0A_{1}\cdots A_{k}a_{1}\cdots a_{n-k}}\rightarrow \varepsilon
^{c_{1}\cdots c_{n}}\varepsilon _{d_{1}\cdots d_{n}}
\]
for the invariant tensor. Various combinations of Kronecker deltas are also
possible for $Q_{A_{1}\cdots A_{m}a_{1}\cdots a_{n-k}}^{B_{1}\cdots
B_{m}b_{1}\cdots b_{n-k}}$.

A scale-invariant dual action of Yang-Mills type is given by
\[
S_{dual}=\int \mathbf{\Omega }_{\tilde{B}}^{\tilde{A}\;\ast }\mathbf{\Omega }%
_{\tilde{A}}^{\tilde{B}}
\]
The resulting field equation, however, is more complicated than the usual $%
\mathbf{D}^{\ast }\mathbf{\Omega }_{\tilde{B}}^{\tilde{A}}=0$, because $%
\delta \mathbf{\omega }_{\tilde{B}}^{\tilde{A}}$\ does not commute with the
dual operator.

The most general linear Lorentz and scale-invariant (weight zero) action
built out of biconformal curvatures and the two invariants $\mathbf{\omega }%
^{a}\mathbf{\omega }_{a}$\ and $\mathbf{d\omega }_{0}^{0}$\ in a $2n$-dim
biconformal space spanned by $\{\omega ^{a_{i}},\omega _{a_{i}};i=1...n\}$\
is a linear combination of $S_{1,0},S_{0,1}$\ and $S_{0,0},$%
\[
S=\int (\alpha \mathbf{\Omega }_{b_{1}}^{a_{1}}+\beta \delta _{b_{1}}^{a_{1}}%
\mathbf{\Omega }_{0}^{0}+\gamma \mathbf{\omega }^{a_{1}}\mathbf{\omega }%
_{b_{1}})\mathbf{\omega }^{a_{2}...a_{n}}\mathbf{\omega }_{b_{2}...b_{n}}%
\varepsilon ^{b_{1}...b_{n}}\varepsilon _{a_{1}...a_{n}}
\]
where $\alpha ,\beta ,\gamma \in \mathbf{R}$\ are constants. Notice that $%
\gamma $ is an arbitrary cosmological constant. An additional term
containing $\mathbf{d\omega }_{0}^{0}$\ would be redundant because of
structure equation (\ref{Structure 4}) for $\Omega _{0}^{0}$. Moreover, $S$\
cannot contain torsion or co-torsion terms, nor is anything further found by
using Kronecker deltas in place of the Levi-Civita tensors. This action and
the resulting field equations will be considered in detail in the following
sections.

Finally, to build $O(n,2)$\ invariants, we return to the full $O(n,2)$\
notation. We can write
\[
S_{n}=\int \mathbf{\Omega }_{\tilde{B}}^{\tilde{A}}\ldots \mathbf{\Omega }_{%
\tilde{D}}^{\tilde{C}}Q_{\tilde{A}\cdots \tilde{C}}^{\tilde{B}\cdots \tilde{D%
}}
\]
where $Q_{\tilde{A}\cdots \tilde{C}}^{\tilde{B}\cdots \tilde{D}}$\ is an $%
O(n,2)$-invariant tensor. $Q_{\tilde{A}\cdots \tilde{C}}^{\tilde{B}\cdots
\tilde{D}}$\ must be built from $\delta _{\tilde{B}}^{\tilde{A}},\eta _{%
\tilde{A}\tilde{B}}$\ or the $(n+2)$-dimensional Levi-Civita tensor. The
only object with the correct index structure is
\[
Q_{\tilde{A}\cdots \tilde{C}}^{\tilde{B}\cdots \tilde{D}}=\delta _{\tilde{A}%
\cdots \tilde{C}}^{\tilde{B}\cdots \tilde{D}}=\frac{1}{2!(n-2)!}\varepsilon
^{\tilde{B}\cdots \tilde{D}\tilde{E}\tilde{F}}\varepsilon _{\tilde{A}\cdots
\tilde{C}\tilde{E}\tilde{F}}
\]
With this specification for $Q_{\tilde{A}\cdots \tilde{C}}^{\tilde{B}\cdots
\tilde{D}},$\ $S_{n}$\ becomes the $n^{th}$\ Pontrijagin class.

\section{The Linear Action}

As noted above, in a $2n$-dim biconformal space the most general Lorentz and
scale-invariant action which is linear in the biconformal curvatures and
structural invariants is
\begin{equation}
S=\int (\alpha \mathbf{\Omega }_{b_{1}}^{a_{1}}+\beta \delta _{b_{1}}^{a_{1}}%
\mathbf{\Omega }_{0}^{0}+\gamma \mathbf{\omega }^{a_{1}}\mathbf{\omega }
_{b_{1}})\mathbf{\omega }^{a_{2}...a_{n}}\mathbf{\omega }_{b_{2}...b_{n}}%
\varepsilon ^{b_{1}...b_{n}}\varepsilon _{a_{1}...a_{n}}  \label{Action}
\end{equation}
We will always assume non-vanishing $\alpha $\ and $\beta $. Variation of
this action with respect to the connection one-forms yields the following
field equations:
\begin{eqnarray}
\delta _{\mathbf{\omega }_{0}^{0}}S=0\Rightarrow 0 &=&\beta (\Omega
^{a}{}_{ba}-2\Omega _{ca}^{\ d}\delta _{db}^{ca})  \label{F-Eq 1a} \\
0 &=&\beta (\Omega _{a}^{\ ba}-2\Omega _{\ a}^{cd}\delta _{dc}^{ab})
\label{F-Eq 1b} \\
\delta _{\mathbf{\omega }_{b}^{a}}S=0\Rightarrow 0 &=&\alpha (-\Delta
_{eg}^{af}\Omega _{\ ab}^{b}+2\Delta _{eb}^{cf}\delta _{dg}^{ab}\Omega
_{ac}^{\ d})  \label{F-Eq 2a} \\
0 &=&\alpha (-\Delta _{eb}^{gf}\Omega _{a}^{\ ab}+2\Delta _{ed}^{af}\delta
_{ab}^{gc}\Omega _{\ c}^{bd})  \label{F-Eq 2b} \\
\delta _{\mathbf{\omega }^{a}}S=0\Rightarrow 0 &=&\alpha \Omega
_{bac}^{a}+\beta \Omega _{0bc}^{0}  \label{F-Eq 3a} \\
0 &=&2(\alpha \Omega _{cd}^{ac}+\beta \Omega _{0d}^{0a})\delta
_{ab}^{ed}+(\alpha (n-1)-\beta +\gamma n^{2})\delta _{b}^{e}  \label{F-Eq 3b}
\\
\delta _{\mathbf{\omega }_{a}}S=0\Rightarrow 0 &=&\alpha \Omega
_{a}^{bac}+\beta \Omega _{0}^{0bc}  \label{F-Eq 4a} \\
0 &=&2(\alpha \Omega _{dc}^{ca}+\beta \Omega _{0d}^{0a})\delta
_{ab}^{ed}+(\alpha (n-1)-\beta +\gamma n^{2})\delta _{b}^{e}  \label{F-Eq 4b}
\end{eqnarray}
Combining equations (\ref{F-Eq 3b}) and (\ref{F-Eq 4b}) we see that the
latter can be replaced by
\begin{equation}
\Omega _{cd}^{ac}=\Omega _{dc}^{ca}  \label{F-Eq 4c}
\end{equation}

\section{Solution for the Curvatures}

We now find the most general solution to these equations subject only to a
constraint of minimal torsion. Starting with the most general ansatz for the
spin connection and Weyl vector, we obtain expressions for the torsion and
co-torsion. Then we find the form of the connection required to satisfy eqs.(%
\ref{F-Eq 1a})-(\ref{F-Eq 2b}). The result does not permit vanishing torsion
without vanishing Weyl vector, so we choose the minimal torsion constraint
consistent with a general form for the Weyl vector. The constraint and field
equations lead to a foliation by $n$-dim flat Riemannian manifolds, possibly
with torsion. By invoking the gauge freedom on each of these manifolds, we
show the existence of a second foliation by $n$-dim Riemannian spacetimes
\textit{without} torsion satisfying the vacuum Einstein equations.
Generically, the full biconformal space also has a symplectic structure.

We first write the spin connection $\mathbf{\omega }_{b}^{a}$ as
\begin{eqnarray*}
\mathbf{\omega }_{b}^{a} &=&\mathbf{\alpha }_{b}^{a}+\mathbf{\beta }_{b}^{a}+%
\mathbf{\gamma }_{b}^{a} \\
&=&(\alpha _{bc}^{a}+\beta _{bc}^{a}+\gamma _{bc}^{a})\mathbf{\omega }
^{c}+(\alpha _{b}^{ac}+\beta _{b}^{ac}+\gamma _{b}^{ac})\mathbf{\omega }_{c}
\end{eqnarray*}
with $\mathbf{\alpha }_{b}^{a}$ and $\mathbf{\beta }_{b}^{a}$ defined by
\begin{eqnarray}
\mathbf{d\omega }^{a} &=&\mathbf{\omega }^{b}\mathbf{\alpha }_{b}^{a}+\frac{1%
}{2}\Omega ^{abc}\mathbf{\omega }_{bc}  \label{Ansatz 1} \\
\mathbf{d\omega }_{a} &=&\mathbf{\beta }_{a}^{b}\mathbf{\omega }_{b}+\frac{1%
}{2}\Omega _{abc}\mathbf{\omega }^{bc}  \label{Ansatz 2}
\end{eqnarray}
Using this ansatz as well as the expanded form of the Weyl vector
\[
\mathbf{\omega }_{0}^{0}=W_{a}\mathbf{\omega }^{a}+W^{a}\mathbf{\omega }_{a}
\]
in structure equations (\ref{Structure 2}) and (\ref{Structure 3}), $\Omega
^{abc}$ and $\Omega _{abc}$ remain related to derivatives of the solder- and
co-solder forms, whereas the other torsion and co-torsion terms are
algebraic in $\mathbf{\alpha }_{b}^{a},\mathbf{\beta }_{b}^{a}$ and $\mathbf{%
\gamma }_{b}^{a}$:
\begin{eqnarray}
\Omega _{\ bc}^{a} &=&\gamma _{cb}^{a}-\gamma _{bc}^{a}+\beta
_{cb}^{a}-\beta _{bc}^{a}+W_{c}\delta _{b}^{a}-W_{b}\delta _{c}^{a}
\label{ST Torsion} \\
\Omega _{\ b}^{ac\ } &=&\gamma _{b}^{ac}+\beta _{b}^{ac}-W^{c}\delta _{b}^{a}
\label{CT Torsion} \\
\Omega _{ac\ }^{\ b} &=&\alpha _{ac}^{b}+\gamma _{ac}^{b}-W_{c}\delta
_{a}^{b}  \label{CT Co-Torsion} \\
\Omega _{a}^{\ bc} &=&\alpha _{a}^{bc}-\alpha _{a}^{cb}+\gamma
_{a}^{bc}-\gamma _{a}^{cb}+W^{b}\delta _{a}^{c}-W^{c}\delta _{a}^{b}
\label{M Co-Torsion}
\end{eqnarray}
Thus, the separation of the connection allows us to solve the first four
field equations algebraically. Imposing field equations (\ref{F-Eq 1a}) and (%
\ref{F-Eq 1b}) onto (\ref{CT Torsion}) and (\ref{CT Co-Torsion}) we get
\begin{eqnarray*}
\beta _{ba}^{a} &=&\alpha _{ba}^{a} \\
\beta _{a}^{ba} &=&\alpha _{a}^{ba}
\end{eqnarray*}
Using this result and imposing field equations (\ref{F-Eq 2a}) and (\ref
{F-Eq 2b}) onto (\ref{ST Torsion}) and (\ref{M Co-Torsion}) completely
determines $\mathbf{\gamma }_{b}^{a}$ in terms of $\mathbf{\alpha }_{b}^{a}$
and $\mathbf{\beta }_{b}^{a}$:
\[
\mathbf{\gamma }_{b}^{a}=-(\alpha _{bc}^{a}\mathbf{\omega }^{c}+\beta
_{b}^{ac}\mathbf{\omega }_{c})
\]
so the spin connection becomes
\[
\mathbf{\omega }_{b}^{a}=\beta _{bc}^{a}\mathbf{\omega }^{c}+\alpha _{b}^{ac}%
\mathbf{\omega }_{c}
\]
Defining the traceless Lorentz tensor
\begin{eqnarray*}
\mathbf{\sigma }_{b}^{a} &\equiv &\mathbf{\alpha }_{b}^{a}-\mathbf{\beta }%
_{b}^{a} \\
&=&\sigma _{bc}^{a}\mathbf{\omega }^{c}+\sigma _{b}^{ac}\mathbf{\omega }_{c}
\end{eqnarray*}
equations (\ref{ST Torsion})-(\ref{M Co-Torsion}) become
\begin{eqnarray}
\Omega _{\ bc}^{a} &=&\sigma _{bc}^{a}-\sigma _{cb}^{a}+W_{c}\delta
_{b}^{a}-W_{b}\delta _{c}^{a}  \label{ST Torsion A} \\
\Omega _{\ b}^{ac\ } &=&-W^{c}\delta _{b}^{a}  \label{CT Torsion A} \\
\Omega _{ac\ }^{\ b} &=&-W_{c}\delta _{a}^{b}  \label{CT Co-Torsion A} \\
\Omega _{a}^{\ bc} &=&\sigma _{a}^{bc}-\sigma _{a}^{cb}+W^{b}\delta
_{a}^{c}-W^{c}\delta _{a}^{b}  \label{M Co-Torsion A}
\end{eqnarray}
While it might seem natural to demand vanishing torsion, $\mathbf{\Omega }%
^{a}=0$, notice that the traces of $\Omega _{\ bc}^{a}$ and $\Omega _{\
b}^{ac}$ are given by
\begin{eqnarray*}
\Omega _{\ ba}^{b} &=&(n-1)W_{a} \\
\Omega _{\ a}^{ac} &=&-nW^{c}
\end{eqnarray*}
Therefore, constraining either eq.(\ref{ST Torsion A}) or eq.(\ref{CT
Torsion A}) to vanish unnecessarily constrains the Weyl vector and thus
greatly reduces the set of allowed geometries. Instead, we impose the
strongest torsion constraint that is consistent with a general Weyl vector,
namely,
\begin{equation}
\mathbf{\Omega }^{a}=\mathbf{\omega }^{a}\mathbf{\omega }_{0}^{0}
\label{Torsion}
\end{equation}
As a consequence,
\begin{eqnarray*}
\Omega ^{abc} &=&0 \\
\sigma _{bc}^{a} &=&\sigma _{cb}^{a}
\end{eqnarray*}
Since the antisymmetry of the spin connection,
\[
\mathbf{\omega }_{b}^{a}=-\eta ^{ac}\eta _{bd}\mathbf{\omega }_{c}^{d}
\]
is inherited by $\mathbf{\alpha }_{b}^{a}$, $\mathbf{\beta }_{b}^{a}$, and
therefore $\mathbf{\sigma }_{b}^{a}$, it is now easy to show by permuting
the indices of $\eta _{ad}\sigma _{bc}^{d}$ and taking the usual linear
combination, that in fact
\[
\sigma _{bc}^{a}\equiv \alpha _{bc}^{a}-\beta _{bc}^{a}=0
\]
so that
\[
\mathbf{\omega }_{b}^{a}=\mathbf{\alpha }_{b}^{a}
\]
We make no assumption concerning the co-torsion, curvature or dilation. In
particular, for the co-torsion we see from eqs.(\ref{CT Co-Torsion A}) and (%
\ref{M Co-Torsion A}) that constraints on $\Omega _{a}^{\ bc}$ or $\Omega
_{ac}^{\ b}$ would constrain $\mathbf{\omega }_{0}^{0}.$ Furthermore, we
will see below that vanishing spacetime co-torsion, $\Omega _{abc}=0$, leads
to vanishing spacetime curvature, $\Omega _{bcd}^{a}=0$, and would therefore
be too strong an assumption.

The torsion constraint makes it possible to obtain an algebraic condition on
the curvatures from the Bianchi identity associated with eq.(\ref{Structure
2}). Taking the exterior derivative of eq.(\ref{Structure 2}) gives
\[
\mathbf{D\Omega }^{a}\equiv \mathbf{d\Omega }^{a}+\mathbf{\Omega }^{b}%
\mathbf{\omega }_{b}^{a}-\mathbf{\omega }_{0}^{0}\mathbf{\Omega }^{a}=%
\mathbf{\omega }^{b}\mathbf{\Omega }_{b}^{a}-\mathbf{\omega }^{a}\mathbf{%
\Omega }_{0}^{0}
\]
Simplifying $\mathbf{D\Omega }^{a}$ using eqs.(\ref{Torsion}) and (\ref
{Structure 4})

\begin{eqnarray*}
\mathbf{D\Omega }^{a} &=&\mathbf{D}(\mathbf{\omega }^{a}\mathbf{\omega }%
_{0}^{0}) \\
&=&-\mathbf{\omega }^{a}\mathbf{\Omega }_{0}^{0}-\mathbf{\omega }^{ab}%
\mathbf{\omega }_{b}
\end{eqnarray*}
(recall that $\mathbf{\omega }^{ab}\equiv \mathbf{\omega }^{a}\mathbf{\omega
}^{b}$) the Bianchi identity reduces to
\[
\mathbf{\omega }^{b}\mathbf{\Omega }_{b}^{a}=-\mathbf{\omega }^{ab}\mathbf{%
\omega }_{b}
\]
which implies
\begin{eqnarray}
\Omega _{\lbrack bcd]}^{a} &=&0  \label{Curvature ST} \\
\Omega _{bd}^{ac} &=&-\Delta _{db}^{ac}  \label{Curvature CT} \\
\Omega _{b}^{acd} &=&0  \label{Curvature MT}
\end{eqnarray}
If we define
\begin{eqnarray*}
\mathbf{R}_{b}^{a} &\equiv &\mathbf{d\omega }_{b}^{a}-\mathbf{\omega }%
_{b}^{c}\mathbf{\omega }_{c}^{a} \\
&=&\mathbf{d\alpha }_{b}^{a}-\mathbf{\alpha }_{b}^{c}\mathbf{\alpha }_{c}^{a}
\end{eqnarray*}
so that $\mathbf{\Omega }_{b}^{a}=\mathbf{R}_{b}^{a}-\Delta _{db}^{ac}%
\mathbf{\omega }_{c}\mathbf{\omega }^{d},$ eqs.(\ref{Curvature CT}) and (\ref
{Curvature MT}) show that $\mathbf{R}_{b}^{a}$ has vanishing cross- and
momentum-terms:
\[
\mathbf{R}_{b}^{a}=\frac{1}{2}R_{bcd}^{a}\mathbf{\omega }^{cd}
\]
This result also follows from the Bianchi identity arising from eq.(\ref
{Ansatz 1}) with $\Omega ^{abc}=0$. Noting from the trace of $\Omega
_{\lbrack bcd]}^{a}=0$ and the antisymmetry condition $\Omega
_{bcd}^{a}=-\eta _{be}\eta ^{af}\Omega _{fcd}^{e}$ that $R_{bac}^{a}\equiv
R_{bc}=R_{cb}$, field equation (\ref{F-Eq 3a}) implies separate vanishing of
the $\alpha $ and $\beta $ terms
\begin{eqnarray}
R_{bc} &=&0  \label{Vacuum Einstein} \\
\Omega _{0bc}^{0} &=&0  \label{ST Dilation}
\end{eqnarray}
while eq.(\ref{F-Eq 4a}) immediately gives
\begin{equation}
\Omega _{0}^{0cd}=0  \label{MT Dilation}
\end{equation}
Notice that, while eq.(\ref{Vacuum Einstein}) is certainly similar to the
vacuum Einstein equation, $\mathbf{R}_{b}^{a}$ has not yet been shown to be
the curvature of a Riemannian geometry. In particular, though it has the
general form of an $n$-dim curvature tensor, it might in principle depend on
all $2n$ coordinates, on the torsion, and/or on the Weyl vector.

Continuing with the field equations, we see that since $\Delta
_{ac}^{ab}=\Delta _{ca}^{ba}=(n-1)\delta _{c}^{b}$, eq.(\ref{F-Eq 4c}) is
identically satisfied. Finally, eq.(\ref{F-Eq 3b}) implies
\begin{equation}
\Omega _{0b}^{0a}=\lambda \delta _{b}^{a}  \label{CT Dilation}
\end{equation}
where
\[
\lambda \equiv \frac{\alpha n(n-1)-\beta +\gamma n^{2}}{\beta (n-1)}
\]
Thus, the entire 3-parameter class of actions leads to a 1-parameter class
of solutions. In particular, the form of the solution is largely independent
of the value of the cosmological constant.

We have now satisfied all of the field equations. The curvatures take the
form
\begin{eqnarray}
\mathbf{\Omega }_{b}^{a} &=&\mathbf{R}_{b}^{a}-\Delta _{db}^{ac}\mathbf{%
\omega }_{c}\mathbf{\omega }^{d}  \label{Curvature} \\
\mathbf{\Omega }_{0}^{0} &=&\lambda \mathbf{\omega }_{a}\mathbf{\omega }^{a}%
\mathbf{\ }  \label{Dilation} \\
\mathbf{\Omega }^{a} &=&\mathbf{\omega }^{a}\mathbf{\omega }_{0}^{0}
\label{Torsion again} \\
\mathbf{\Omega }_{a} &=&\mathbf{\omega }_{0}^{0}\mathbf{\omega }_{a}+\sigma
_{a}^{bc}\mathbf{\omega }_{bc}+\frac{1}{2}\Omega _{abc}\mathbf{\omega }^{bc}
\label{Co-Torsion}
\end{eqnarray}
with
\begin{eqnarray*}
R_{ab} &=&0 \\
\sigma _{a}^{ba} &=&0
\end{eqnarray*}
In the next section, we find further constraints on the curvatures arising
from the structure equations. We also find an explicit form for the
connection that displays clearly the minimal field content of the general
solution.

\section{Solution for the Connection}

While eqs.(\ref{Curvature})-(\ref{Co-Torsion}) for the curvatures satisfy
all of the field equations, they do not fully incorporate the form of the
biconformal structure equations as embodied in the Bianchi identities.
Moreover, it is not yet clear what minimal field content is required to
insure a unique solution. Therefore, in this section, we turn to the
consequences of the form (\ref{Curvature})-(\ref{Co-Torsion}) of the
curvatures on the connection.

Substituting the reduced curvatures into eqs.(\ref{Structure 1})-(\ref
{Structure 4}), the structure equations now take the form
\begin{eqnarray}
\mathbf{R}_{b}^{a} &=&\mathbf{d\alpha }_{b}^{a}-\mathbf{\alpha }_{b}^{c}%
\mathbf{\alpha }_{c}^{a} \\
\mathbf{d\omega }^{a} &=&\mathbf{\omega }^{b}\mathbf{\alpha }_{b}^{a}
\label{Structure 2a} \\
\mathbf{d\omega }_{a} &=&\mathbf{\alpha }_{a}^{b}\mathbf{\omega }_{b}-%
\mathbf{\sigma }_{a}^{b}\mathbf{\omega }_{b}+\frac{1}{2}\Omega _{abc}\mathbf{%
\omega }^{bc}  \label{Structure 3a} \\
\mathbf{d\omega }_{0}^{0} &=&(1-\lambda )\mathbf{\omega }^{a}\mathbf{\omega }%
_{a}  \label{Structure 4a}
\end{eqnarray}

We begin with eq.(\ref{Structure 4a}). For $\lambda \neq 0,1$, eqs.(\ref
{Structure 4a}) and (\ref{Dilation}) show that the three biconformal
invariants $\mathbf{d\omega }_{0}^{0}$, $\mathbf{\omega }_{a}\mathbf{\omega }%
^{a}$ and $\mathbf{\Omega }_{0}^{0}$ noted in the introduction are all
proportional to each other. Moreover, eq.(\ref{Structure 4a}) shows in
gestalt form that each invariant is a symplectic form: $\mathbf{d\omega }%
_{0}^{0}$ is manifestly closed and $\mathbf{\omega }_{a}\mathbf{\omega }^{a}$
is manifestly nondegenerate. A generic biconformal space subject to the
linear action and minimal torsion condition $\mathbf{\Omega }^{a}=\mathbf{%
\omega }^{a}\mathbf{\omega }_{0}^{0}$ is therefore a symplectic manifold. By
a well-known theorem \cite{Choquet-Bruhat} it is always possible to
construct an almost complex structure and a K\"{a}hler metric on a
symplectic manifold. Therefore the field equations arising from the linear
action constrain the biconformal space to be almost K\"{a}hler.

If $\lambda =0$, the dilation vanishes, while $\mathbf{\omega }^{a}\mathbf{%
\omega }_{a}=\mathbf{d\omega }_{0}^{0}$ remains symplectic. All subsequent
calculations hold with $\mathbf{\Omega }_{0}^{0}=0$. This case has been
investigated in \cite{New Conformal Gauging Paper}, where it was argued that
for \textit{classical} geometries it is reasonable to assume that no path in
phase space encloses a plaquette on which the dilation is nonvanishing. Such
spaces were shown to be in $1-1$ correspondence with $n$-dimensional
Einstein-Maxwell spacetimes.

We will consider the special case $\lambda =1$ further in Sec.(7). For the
remainder of this Section let $\lambda \neq 1$.

The Bianchi identity obtained by taking the exterior derivative of eq.(\ref
{Structure 4a}) is
\begin{eqnarray*}
0 &=&\mathbf{d\omega }_{a}\mathbf{\omega }^{a}-\mathbf{\omega }_{a}\mathbf{%
d\omega }^{a} \\
&=&(\mathbf{\alpha }_{a}^{b}\mathbf{\omega }_{b}-\mathbf{\sigma }_{a}^{b}%
\mathbf{\omega }_{b}+\frac{1}{2}\Omega _{abc}\mathbf{\omega }^{bc})\mathbf{%
\omega }^{a}-\mathbf{\omega }_{a}\mathbf{\omega }^{b}\mathbf{\alpha }_{b}^{a}
\\
&=&\mathbf{\sigma }_{a}^{b}\mathbf{\omega }^{a}\mathbf{\omega }_{b}+\frac{1}{%
2}\Omega _{abc}\mathbf{\omega }^{abc}
\end{eqnarray*}
so that
\begin{eqnarray*}
\Omega _{\lbrack abc]} &=&0 \\
\mathbf{\sigma }_{a}^{b} &=&0
\end{eqnarray*}
Here the vanishing of $\mathbf{\sigma }_{a}^{b}$ follows from the
simultaneous vertical antisymmetry and horizontal symmetry of $\sigma
_{a}^{bc}.$ This vanishing of $\mathbf{\sigma }_{a}^{b}\mathbf{\ }$amounts
to the vanishing of the momentum term of the co-torsion.

Next we examine consequences of eq.(\ref{Structure 2a}), which is in
involution. By the Frobenius theorem, we can consistently set $\mathbf{%
\omega }^{a}$ to zero and obtain a foliation by submanifolds where the spin
connection and Weyl vector reduce to
\begin{eqnarray*}
\mathbf{\hat{\alpha}}_{b}^{a} &\equiv &\mathbf{\alpha }_{b|\mathbf{\omega }%
^{a}=0}^{a}=\alpha _{b}^{ac}\mathbf{\omega }_{c} \\
\mathbf{\hat{\omega}}_{0}^{0} &\equiv &\mathbf{\omega }_{0|\mathbf{\omega }%
^{a}=0}^{0}=W^{a}\mathbf{\omega }_{a}
\end{eqnarray*}
If we define
\[
\mathbf{f}_{a}\equiv \mathbf{\omega }_{a|\mathbf{\omega }^{a}=0}
\]
then each submanifold is described by the reduced structure equations
\begin{eqnarray*}
\mathbf{d\hat{\alpha}}_{b}^{a} &=&\mathbf{\hat{\alpha}}_{b}^{c}\mathbf{\hat{%
\alpha}}_{c}^{a} \\
\mathbf{df}_{a} &=&\mathbf{\hat{\alpha}}_{a}^{b}\mathbf{f}_{b}+\mathbf{f}_{a}%
\mathbf{\hat{\omega}}_{0}^{0} \\
\mathbf{d\hat{\omega}}_{0}^{0} &=&0
\end{eqnarray*}
Since $\mathbf{\hat{\omega}}_{0}^{0}$\ is closed, we can scale-gauge the
Weyl vector to zero on each subspace, i.e. $W^{a}=0$. The remaining two
equations then describe a flat $n$-dimensional Riemannian spacetime. Since
the spin-connection is involute, there also exists a Lorentz gauge
transformation such that $\mathbf{\hat{\alpha}}_{b}^{a}=0$ on each
submanifold, i.e. $\alpha _{b}^{ac}=0$. With these gauge choices the system
reduces to simply $\mathbf{df}_{a}=0,$ with solution $\mathbf{f}_{a}=\mathbf{%
\ d}\theta _{a}$ for some $0$-forms $\theta _{a}.$

Returning to the full biconformal space, we now have a gauge such that the
spin connection and Weyl vector are
\begin{eqnarray}
\mathbf{\alpha }_{b}^{a} &=&\alpha _{bc}^{a}\mathbf{\omega }^{c}
\label{Spin connection} \\
\mathbf{\omega }_{0}^{0} &=&W_{a}\mathbf{\omega }^{a}  \label{Weyl vector}
\end{eqnarray}
while the co-solder form may be written in terms of $\mathbf{f}_{a}$ and an
additional term linear in the solder form,
\[
\mathbf{\omega }_{a}=\mathbf{f}_{a}+h_{ab}\mathbf{\omega }^{b}
\]
Notice that $\mathbf{f}_{a}$ is essentially unchanged by this extension,
except that the $0$-forms $\theta _{a}$ must be regarded as dependent on all
$2n$ coordinates. This means that $\mathbf{df}_{a}$ remains at least linear
in $\mathbf{f}_{a},$ and is consequently involute (see Appendix B). We can
therefore turn the problem around, setting $\mathbf{f}_{a}=0$ to obtain a
second foliation of the biconformal space. We can define $\mathbf{h}_{a}$ in
terms of this involution, setting
\[
\mathbf{h}_{a}\equiv \mathbf{\omega }_{a|\mathbf{f}_{a}=0}=h_{ab}\mathbf{\
\omega }^{b}
\]
with $h_{ab}$ arbitrary. Now, with $\mathbf{f}_{a}=0,$ the new submanifolds
are described by
\begin{eqnarray*}
\mathbf{R}_{b}^{a} &=&\mathbf{d\alpha }_{b}^{a}-\mathbf{\alpha }_{b}^{c}%
\mathbf{\alpha }_{c}^{a} \\
\mathbf{d\omega }^{a} &=&\mathbf{\omega }^{b}\mathbf{\alpha }_{b}^{a} \\
\mathbf{dh}_{a} &=&\mathbf{\alpha }_{a}^{b}\mathbf{h}_{b}+\frac{1}{2}\Omega
_{abc}\mathbf{\omega }^{bc} \\
\mathbf{d\omega }_{0}^{0} &=&(1-\lambda )\mathbf{\omega }^{a}\mathbf{h}_{a}
\end{eqnarray*}
The first two equations are unchanged from their full biconformal form,
showing that the curvature $\mathbf{R}_{b}^{a}$ and connection $\mathbf{%
\alpha }_{b}^{a}$ (and of course $\mathbf{\omega }^{a}$, by the first
involution) are fully determined on the $\mathbf{f}_{a}=0$ submanifold.
Thus, $\mathbf{\alpha }_{b}^{a}$ is the usual spin connection compatible
with $\mathbf{\omega }^{a},$ while $\mathbf{R}_{b}^{a}$ is its curvature.
Therefore, the vanishing of the Ricci tensor, $R_{ab}=0,$ now shows that
these $n$-dim submanifolds satisfy the vacuum Einstein equations. Even
though the torsion and dilation have nonvanishing spacetime projections, $%
\Omega _{abc}$ and $\Omega _{0b|\mathbf{f}_{a}=0}^{0a}=-\lambda \mathbf{%
\omega }^{a}\mathbf{h}_{a},$ respectively, the curvature is the one computed
from the solder form $\mathbf{\omega }^{a}$ alone; even though our action
included an arbitrary cosmological constant, $\gamma $, the Ricci tensor
vanishes. This is our most important result, since it establishes a direct
connection between the usual Ricci-flat Riemannian structure of general
relativity and the more general structure of conformal gauge theory.

\smallskip

Finally, we seek a minimum set of fields required to uniquely specify a
complete solution. We can easily find such a minimum set by choosing
coordinates. Based on the involution for $\mathbf{\omega }^{a}$ there exist $%
n$ coordinates $x^{\mu }$ such that
\[
\mathbf{\omega }^{a}=e_{\mu }^{\;a}\mathbf{d}x^{\mu }
\]
with the component matrices necessarily invertible. From eq.(\ref{Structure
2a}), we immediately find that $e_{\mu }^{\;a}=e_{\mu }^{\;a}(x).$
Similarly, we show in Appendix B that there exist coordinates $y_{\nu }$
such that $\mathbf{f}_{a}$ takes the form
\[
\mathbf{f}_{a}=e_{a}^{\;\mu }\mathbf{d}y_{\mu }+\psi _{a\mu }\mathbf{d}%
x^{\mu }
\]
where $e_{a}^{\;\mu }$ is the inverse to $e_{\mu }^{\;a}$ and $\psi _{a\mu
}=\psi _{a\mu }(x,y)$.

Using this coordinate choice and writing the co-solder form as
\begin{eqnarray}
\mathbf{\omega }_{a} &=&e_{a}^{\;\mu }(\mathbf{d}y_{\mu }+h_{\mu \nu }%
\mathbf{d}x^{\nu }+\psi _{\mu \nu }\mathbf{d}x^{\nu })  \nonumber \\
&\equiv &e_{a}^{\;\mu }(\mathbf{d}y_{\mu }+\hat{h}_{\mu \nu }\mathbf{d}
x^{\nu })  \label{Co-solder coord}
\end{eqnarray}
eq.(\ref{Structure 4a}) yields
\begin{eqnarray*}
\partial _{[\nu }W_{\mu ]} &=&(\lambda -1)\hat{h}_{[\mu \nu ]} \\
\partial ^{\nu }W_{\mu } &=&(\lambda -1)\delta _{\mu }^{\nu }
\end{eqnarray*}
where $e_{a}^{\;\mu }$ and $e_{\mu }^{\;a}$ are used to interchange
coordinate and orthonormal indices in the usual way and $(\partial _{\nu
},\partial ^{\nu })$ denote derivatives with respect to $(x^{\mu },y_{\nu
}). $ The second equation can immediately be integrated,
\begin{equation}
W_{\mu }=(\lambda -1)(y_{\mu }-A_{\mu }(x))  \label{Integrated Weyl vector}
\end{equation}
where the integration ``constant'' $A_{\mu }(x)$ determines the
antisymmetric part of $\hat{h}_{[\mu \nu ]}:$
\begin{equation}
\hat{h}_{[\mu \nu ]}=\partial _{[\nu }A_{\mu ]}  \label{Curl A}
\end{equation}
Clearly, $\hat{h}_{[\mu \nu ]}$ is a function of $x$ only.

The fields $A_{\mu }$ and $\hat{h}_{[\mu \nu ]}$ in eqs.(\ref{Integrated
Weyl vector}) and (\ref{Curl A}) are purely coordinate dependent. To see
this, note that under coordinate transformations of the form $\bar{y}%
_{\alpha }=y_{\alpha }+$ $\gamma _{\alpha }(x),$ eq.(\ref{Co-solder coord})
changes to
\[
\mathbf{\omega }_{a}=e_{a}^{\;\mu }(\mathbf{d}\bar{y}_{\mu }+(\hat{h}_{\mu
\nu }+\gamma _{\mu ,\nu })\mathbf{d}x^{\nu })
\]
Then eqs.(\ref{Integrated Weyl vector}) and (\ref{Curl A}) become
\begin{eqnarray*}
\bar{W}_{\mu } &=&(\lambda -1)(\bar{y}_{\mu }-\gamma _{\mu }-A_{\mu }(x)) \\
\bar{h}_{[\mu \nu ]} &=&\partial _{[\nu }A_{\mu ]}+\gamma _{[\mu ,\nu ]}
\end{eqnarray*}
Therefore, if we choose $\gamma _{\mu }=-A_{\mu },$ we have simply
\begin{eqnarray*}
\bar{W}_{\mu } &=&(\lambda -1)\hat{y}_{\mu } \\
\bar{h}_{[\mu \nu ]} &=&0
\end{eqnarray*}
We make this coordinate choice below, dropping the overbars.

Finally, with $\mathbf{\sigma }_{a}^{b}=0,$ eq.(\ref{Structure 3a}) for the
co-solder form reduces to
\[
\mathbf{d\omega }_{a}=\mathbf{\alpha }_{a}^{b}\mathbf{\omega }_{b}+\frac{1}{2%
}\Omega _{abc}\mathbf{\omega }^{bc}
\]
or in coordinate form

\begin{eqnarray}
(\partial _{\mu }h_{a\nu }-\alpha _{a\mu }^{b}h_{b\nu })-(\partial _{\nu
}h_{a\mu }-\alpha _{a\nu }^{b}h_{b\mu }) &=&\Omega _{a\mu \nu }  \label{h-x}
\\
\partial ^{\nu }h_{a\mu } &=&\partial _{\mu }e_{a}^{\;\nu }-\alpha
_{ac}^{b}e_{\mu }^{c}e_{b}^{\;\nu }  \label{h-y}
\end{eqnarray}
Using the well-known relation between the orthonormal connection $\mathbf{%
\alpha }_{b}^{a}(x)$ and the Christoffel connection $\Gamma _{\alpha \mu
}^{\nu }(x)$ of a metric compatible geometry,
\[
D_{\mu }e_{a}^{\;\nu }\equiv \partial _{\mu }e_{a}^{\;\nu }-\alpha _{a\mu
}^{b}e_{b}^{\;\nu }+e_{a}^{\;\alpha }\Gamma _{\alpha \mu }^{\nu }=0
\]
we find from eq.(\ref{h-y}) that
\[
\partial ^{\nu }h_{\alpha \mu }=e_{\alpha }^{\;a}\partial ^{\nu }h_{a\mu
}=-\Gamma _{\alpha \mu }^{\nu }
\]
which integrates to
\[
h_{\alpha \mu }=-y_{\nu }\Gamma _{\alpha \mu }^{\nu }+k_{\alpha \mu }(x)
\]
where the symmetric tensor $k_{\alpha \mu }(x)$ is a second integration
``constant''. The contraction with $y_{\mu }$ permits $h_{ab}$ to behave as
a Lorentz tensor. Notice that the covariant curl of $h_{\alpha \mu }$ has a
term proportional to the curvature tensor because the contraction $-y_{\nu
}\Gamma _{\alpha \mu }^{\nu }$ eliminates the extra connection terms that
normally prevent the covariant curl of $\Gamma _{\alpha \mu }^{\nu }$ from
being simply related to curvature. When this result for $h_{\alpha \mu }$ is
substituted into eq.(\ref{h-x}), the spacetime co-torsion is expressed in
terms of $k_{\alpha \mu }(x)$ and the curvature. After restoring the basis
forms,
\[
\frac{1}{2}\Omega _{abc}\mathbf{\omega }^{bc}=-y_{b}\mathbf{R}_{a}^{b}+%
\mathbf{\ Dk}_{a}
\]
where $\mathbf{D}$ is the metric compatible covariant exterior derivative
and $\mathbf{k}_{a}=k_{ab}\mathbf{\omega }^{b}$. As claimed in Sec.($5$),
vanishing $\Omega _{abc}$ implies vanishing curvature since $\mathbf{k}_{a}$
depends on $x^{\mu }$ only.

Collecting the results for the connection, we immediately see the essential
field content:

\begin{eqnarray}
\mathbf{\alpha }_{b}^{a} &=&\mathbf{\alpha }_{b}^{a}(e_{\mu }^{\;a}(x))
\nonumber \\
\mathbf{\omega }^{a} &=&e_{\mu }^{\;a}(x)\mathbf{d}x^{\mu }  \nonumber \\
\mathbf{\omega }_{a} &=&e_{a}^{\;\mu }(x)(\mathbf{d}y_{\mu }-y_{\nu }\Gamma
_{\mu \alpha }^{\nu }\mathbf{d}x^{\alpha }+k_{\mu \alpha }(x)\mathbf{\ d}%
x^{\alpha })  \nonumber \\
&=&\mathbf{D}y_{a}+\mathbf{k}_{a}(x)  \nonumber \\
\mathbf{\omega }_{0}^{0} &=&(\lambda -1)\ y_{\mu }\mathbf{d}x^{\mu }
\label{Final connection}
\end{eqnarray}
The entire solution depends on two fields, $e_{\mu }^{\;a}(x)$ and $k_{\mu
\alpha }=k_{(\mu \alpha )}(x).$ Both of these fields are defined entirely on
the vacuum Einstein spacetime submanifolds (with coordinates $x^{\mu }$).
Otherwise $k_{\mu \alpha }(x)$ is an arbitrary integration constant, while $%
e_{\mu }^{\;a}(x)$ is the usual solder form.

Finally, we write the final form of the curvatures. Decomposing the Riemann
curvature into its traceless and Ricci parts
\begin{equation}
\mathbf{R}_{b}^{a}=\mathbf{C}_{b}^{a}+\Delta _{db}^{ac}\mathcal{R}_{c}%
\mathbf{e}^{d}  \label{Conformal Tensor}
\end{equation}
where
\begin{equation}
\mathcal{R}_{a}\equiv (R_{ab}-\frac{1}{2(n-1)}\eta _{ab}R)\mathbf{e}^{b}
\label{Conformal Ricci}
\end{equation}
we use the Ricci-flat condition together with the results of this section
for $\mathbf{\sigma }_{a}^{b}$ and the co-torsion to write the curvatures as
\begin{eqnarray}
\mathbf{\Omega }_{b}^{a} &=&\mathbf{C}_{b}^{a}-\Delta _{db}^{ac}\mathbf{%
\omega }_{c}\mathbf{\omega }^{d}  \label{Final curvature} \\
\mathbf{\Omega }_{0}^{0} &=&\lambda \mathbf{\omega }_{a}\mathbf{\omega }^{a}%
\mathbf{\ }  \label{Final dilation} \\
\mathbf{\Omega }^{a} &=&\mathbf{\omega }^{a}\mathbf{\omega }_{0}^{0}
\label{Final torsion} \\
\mathbf{\Omega }_{a} &=&\mathbf{\omega }_{0}^{0}\mathbf{\omega }_{a}-y_{b}%
\mathbf{C}_{a}^{b}+\mathbf{Dk}_{a}  \label{Final co-torsion}
\end{eqnarray}

\section{Special Case}

For the special case $\lambda =1$, eq.(\ref{Structure 4a}) implies a closed
and hence locally removable Weyl vector, i.e. $W_{a}=0$. In that case, eqs.(%
\ref{Torsion again}) and (\ref{Co-Torsion}) for the torsion and co-torsion
become
\begin{eqnarray*}
\mathbf{\Omega }^{a} &=&0 \\
\mathbf{\Omega }_{a} &=&\sigma _{a}^{bc}\mathbf{\omega }_{bc}+\frac{1}{2}%
\Omega _{abc}\mathbf{\omega }^{bc}
\end{eqnarray*}
with
\[
\sigma _{a}^{ba}=0
\]
As before, the involution in eq.(\ref{Structure 2a}) allows us to gauge the
spin connection $\mathbf{\alpha }_{b}^{a}$ so that $\alpha _{b}^{ac}=0$.
Then structure equation (\ref{Structure 3}) becomes
\begin{equation}
\mathbf{d\omega }_{a}=\alpha _{ac}^{b}\mathbf{\omega }^{c}\mathbf{\omega }%
_{b}-\sigma _{a}^{bc}\mathbf{\omega }_{bc}+\frac{1}{2}\Omega _{abc}\mathbf{%
\omega }^{bc}  \label{Structure 3b}
\end{equation}
Writing the co-solder form again as
\[
\mathbf{\omega }_{a}=\mathbf{f}_{a}+\mathbf{h}_{a}
\]
such that $\mathbf{f}_{a}\equiv \mathbf{\omega }_{a|\mathbf{\omega }^{a}=0}$%
, we have on the $\mathbf{\omega }^{a}=0$ subspace
\[
\mathbf{df}_{a}=\sigma _{a}^{bc}\mathbf{f}_{bc}
\]
This can be solved in the usual way giving $\sigma _{a}^{bc}$ in terms of
the projected part of $\mathbf{f}_{a}$ and its $y$ derivatives. Since this
solution has the same form at each point $x,$ the expression for $\sigma
_{a}^{bc}$ remains valid when the $x$-dependence of $\mathbf{f}_{a}$ is
restored.

For the $\mathbf{f}_{a}=0$ subspace,
\[
\mathbf{dh}_{a}=\alpha _{ac}^{b}\mathbf{\omega }^{c}\mathbf{h}_{b}-\sigma
_{a}^{bc}\mathbf{h}_{bc}+\frac{1}{2}\Omega _{abc}\mathbf{\omega }^{bc}
\]
In general, this equation determines the spacetime co-torsion, $\Omega
_{abc},$ once $\mathbf{h}_{a}$ is given.

Extending back to the full space, and introducing coordinates as before, we
write
\begin{eqnarray*}
\mathbf{df}_{a} &=&\partial _{\alpha }f_{a}^{\;\beta }\mathbf{d}x^{\alpha }%
\mathbf{d}y_{\beta }+\partial ^{\alpha }f_{a}^{\;\beta }\mathbf{d}y_{\alpha }%
\mathbf{d}y_{\beta } \\
\mathbf{dh}_{a} &=&\partial _{\alpha }h_{a\beta }\mathbf{d}x^{\alpha }%
\mathbf{d}x^{\beta }+\partial ^{\alpha }h_{a\beta }\mathbf{d}y_{\alpha }%
\mathbf{d}x^{\beta }
\end{eqnarray*}
where we can no longer restrict the functional dependence of $f_{a}^{\;\beta
}.$ Now eq.(\ref{Structure 3b}) implies
\[
\partial _{\alpha }f_{a}^{\;\beta }-\partial ^{\beta }h_{a\alpha }=\alpha
_{a\alpha }^{b}f_{b}^{\;\beta }-(\sigma _{a}^{bc}-\sigma
_{a}^{cb})h_{b\alpha }f_{c}^{\;\beta }
\]
Any solution of this equation for $f_{a}^{\;\beta }$ and $h_{a\alpha }$
gives a complete solution. It is clear that solutions do exist, since the $%
\lambda \neq 1$ condition $\sigma _{a}^{bc}=0$ permits the generic solution
to hold.

\section{Comparisons with previous theories}

As mentioned in the introduction, there have been a number of studies of
conformal and superconformal gauge theories for $n>2$. In this section, we
compare our results with these other approaches. The gravitational sectors
of standard conformal actions fall into three principal types:

\begin{enumerate}
\item  Chern-Simons action ($n=3$)

\item  Curvature-linear action with compensating fields ($n=4$)

\item  Curvature-quadratic action with compensating fields ($n\geq 4$)
\end{enumerate}

We will treat each of these cases in turn, comparing the results to ours.

\subsection{Chern-Simons action}

The topological Chern-Simons action, which is intrinsically odd-dim,
is of particular interest in $3$-dimensional conformal gravity \cite
{Deser+Jackiw+Templeton}, where it becomes
\begin{equation}
S=\int \mathbf{\omega }_{b}^{a}(\mathbf{d\omega }_{a}^{b}-\frac{2}{3}\mathbf{%
\omega }_{a}^{c}\mathbf{\omega }_{c}^{b}\mathbf{)}  \label{Chern Simons}
\end{equation}
Here $\mathbf{\omega }_{b}^{a}$ is the spin connection of a torsion-free
Riemannian geometry. When this action is varied with respect to the solder
form $\mathbf{e}^{a}$, the resulting field equation is
\begin{equation}
\mathbf{D}\mathcal{R}_{a}=0  \label{Curl of conformal Ricci}
\end{equation}
with $\mathcal{R}_{a}$ defined as in eq.(\ref{Conformal Ricci}). This is
precisely the condition for spacetime to be conformally flat in $3$-dim \cite
{Eisenhart}, so the model is exactly soluble with $\mathbf{e}^{a}=e^{\phi }%
\mathbf{d}x^{a}$ for any function $\phi (x)$. It has been observed that this
action can be derived from the Chern-Simons action for the whole
(super)conformal group $O(3,2)$ by imposing the constraints of vanishing
curvature and torsion \cite{van Nieuwenhuizen}. The same result follows
\textit{without} constraints if one replaces the Riemannian spin connection
in eq.(\ref{Chern Simons}) by the conformal connection,
\[
\mathbf{\omega }_{b}^{a}\rightarrow \mathbf{\omega }_{B}^{A}
\]
and performs a Palatini variation \cite{Witten}. Then all conformal
curvatures vanish and gauging the Weyl vector to zero renders the first- and
second-order formalisms equivalent again giving condition (\ref{Curl of
conformal Ricci}) for conformal flatness. As in the quadratic $4$-dim theory
(see below), the special conformal gauge field is found to be equal to $%
\mathcal{R}_{a}$.

In biconformal space, the $3$-dim example does not lead to many
simplifications over the general method of solution, though an explicit
check in that dimension did confirm our previous results. A generalization
of condition (\ref{Curl of conformal Ricci}) obviously arises in this case\
(and in fact for $n>3$ as well), since our solution shows the existence of $3
$-dimensional embedded Ricci-flat spacetimes. While our proof demonstrates
the existence of the appropriate gauge choice directly, it is clear that any
other $x$-dependent gauge transformation must lead to a slicing satisfying
eq.(\ref{Curl of conformal Ricci}). In addition, the biconformal model
permits $y$-dependent gauge choices. Thus, the field equations of the linear
biconformal field theory generalize eq.(\ref{Curl of conformal Ricci}) to an
embedding biconformal or phase space. Note further that constraint (\ref
{Constraint 1}) follows from the field equations as in \cite{Witten} rather
than being imposed as in \cite{van Nieuwenhuizen}.

\subsection{Curvature-linear actions}

In standard $4$-dimensional Weyl gauge theory \cite{Freund}-\cite{Bicknell},
one obtains a Lorentz- and scale invariant linear action through the
introduction of a Brans-Dicke-like \cite{Brans-Dicke} compensating field $%
\phi (x)$ in the manner suggested by Deser \cite{Deser} and Dirac \cite
{Dirac}. In close analogy to the geometrical gauge approach of identifying
the gauge fields with connections on spacetime developed by Utiyama and
Kibble \cite
{Utiyama/Kibble} for Poincar\'{e} gravity, a Weyl-covariant derivative $%
D_{a} $ is built out of the spin connection (usually assumed to be metric
compatible and torsion-free) and the Weyl vector. The free (vacuum) action
comprises a kinetic term $\phi \fbox{} \phi $, a Yang-Mills-type term $%
F_{ab}F^{ab}$ for the curl of the Weyl vector $F_{ab}$=$D_{[a}W_{b]}$, a
gravitational term $R\phi ^{2}$, and possibly a cosmological term $\Lambda
\phi ^{4}$:
\begin{equation}
S=\int \sqrt{-g}(6\phi \fbox{} \phi +\frac{1}{4}F_{ab}F^{ab}-R\phi
^{2}+\Lambda \phi ^{4})d^{4}x  \label{Dirac Action}
\end{equation}
Dropping the cosmological term, the corresponding gravitational field
equations change the vacuum Einstein equations \cite{Kasuya} to the
``generalized'' Einstein equations:
\[
2\phi ^{2}(R_{b}^{a}-\frac{1}{2}\delta _{b}^{a}R)+4(D^{a}\phi D_{b}\phi -%
\frac{1}{2}\delta _{b}^{a}D^{c}\phi D_{c}\phi )=T_{b}^{a}
\]
where $T_{b}^{a}$ is the generalized Maxwell stress tensor:
\[
T_{b}^{a}=F_{ac}F^{bc}-\frac{1}{4}\delta _{a}^{b}F_{cd}F^{cd}
\]
In the Einstein gauge one sets $\phi =1$, so the vacuum Einstein equations,
coupled to a spin-1 field, are recovered. However, note that the geometric
meaning of $F_{\mu \nu }$ as producing changes in the lengths of transported
vectors precludes interpreting $F_{\mu \nu }$ as the Maxwell field \cite
{Einstein}.

It is also possible to couple $n$-dimensional \textit{conformal} gravity to
compensating fields of conformal weight $-(n-2)/2$ \cite{Bergshoef}. This
approach does not require an explicit gravity term in the action, since the
d'Alembertian is built out of derivatives that are also covariant with
respect to special conformal transformations and hence contain the special
conformal gauge field $f_{b}^{a}$. Then the Lagrangian $\phi \fbox{} \phi $,
when broken up based on its conformal invariance properties, contains a term
of the form $f_{a}^{a}\phi ^{2}$, which under the conventional constraint (%
\ref{Constraint 1}) reduces to $R\phi ^{2}$ when the Weyl vector is gauged
to zero. As in the Weyl case, this theory is equivalent to Einstein gravity
when $\phi $ is expressed in a particular gauge using the special conformal
gauge freedom.

Biconformal space improves on these results in two important ways: (1)
biconformal space does not require compensating fields, and (2) the first
conventional constraint, eq.(\ref{Constraint 1}) follows from the field
equations and is not required as a constraint. A third point developed
elsewhere \cite{New Conformal Gauging Paper} is that it is possible to
include electromagnetism without the usual interpretational difficulties.

There are interesting differences between these treatments and our results
regarding the effect of constraint (\ref{Constraint 2}). In standard
conformal gaugings vanishing torsion leads to vanishing Weyl vector as a
possible gauge choice and identification of the special conformal gauge
field with $\mathcal{R}_{a}$:
\begin{equation}
f_{a}=-\left( \frac{1}{n-2}\right) \mathcal{R}_{a}
\label{Special conformal is curly R}
\end{equation}
We find that the same results occur if the biconformal torsion is set to
zero and attention is restricted to the $y=0$ subspace. However, on the full
biconformal space, where $y$ is allowed to vary, this solution proves to be
inconsistent. Instead, the torsion may be fixed intrisically by the minimal
torsion constraint, eq.(\ref{Torsion}), resulting in a non-trivial $y$%
-dependance for the Weyl vector and independence of the projected co-solder
form, $k_{a\alpha }$.

To further compare these standard results to the biconformal solution,
consider the final form of the biconformal curvatures
\begin{eqnarray}
\mathbf{\Omega }_{b}^{a} &=&\mathbf{C}_{b}^{a}-\Delta _{db}^{ac}\mathbf{%
\omega }_{c}\mathbf{\omega }^{d} \\
\mathbf{\Omega }_{0}^{0} &=&\lambda \mathbf{\omega }_{a}\mathbf{\omega }^{a}%
\mathbf{\ } \\
\mathbf{\Omega }^{a} &=&\mathbf{\omega }^{a}\mathbf{\omega }_{0}^{0} \\
\mathbf{\Omega }_{a} &=&\mathbf{\omega }_{0}^{0}\mathbf{\omega }_{a}-y_{b}%
\mathbf{C}_{a}^{b}+\mathbf{Dk}_{a}
\end{eqnarray}
The first constraint, eq.(\ref{Constraint 1}), already holds for the \textit{%
spacetime} components of $\mathbf{\Omega }_{b}^{a},$ namely, $\mathbf{\Omega
}_{bcd}^{a}=C_{bcd}^{a}.$ In the standard conformal gauging, this constraint
also includes the term $-\Delta _{db}^{ac}\mathbf{\omega }_{c}\mathbf{\omega
}^{d},$ so that the constraint fixes $\mathbf{\omega }_{c}.$ However, in
biconformal space, $\mathbf{\omega }_{c}$ is independent of $\mathbf{\omega }%
^{d},$ and even its $y=const.$ projection, $\mathbf{k}_{a},$ remains
arbitrary. If we restrict attention to the $y=0$ submanifold, the curvatures
take the form

\begin{eqnarray*}
\mathbf{\Omega }_{b}^{a} &=&\mathbf{C}_{b}^{a}-\Delta _{db}^{ac}\mathbf{k}%
_{c}\mathbf{e}^{d} \\
\mathbf{\Omega }_{0}^{0} &=&0 \\
\mathbf{\Omega }^{a} &=&0 \\
\mathbf{\Omega }_{a} &=&\mathbf{Dk}_{a}
\end{eqnarray*}
It is amusing to notice that if we demand that $\mathbf{\Omega }_{b}^{a}$ be
the Riemannian curvature of the submanifold, then not only do we immediately
have
\begin{equation}
\mathbf{k}_{a}=-\left( \frac{1}{n-2}\right) \mathcal{R}_{a}
\end{equation}
but also the second Bianchi identity $\mathbf{D\Omega }_{b}^{a}=0$ implies $%
\mathbf{D}\mathcal{R}_{a}=0,$ i.e., the spacetime is conformally Ricci flat.
However, $\mathbf{\Omega }_{b}^{a}$ is \textit{not} a Riemannian curvature,
and satisfies a different Bianchi identity that leaves $\mathbf{k}_{a}$
arbitrary. These same comments apply when $n=3$ by simply setting $\mathbf{C}%
_{b}^{a}=0.$

\subsection{Curvature-quadratic actions}

In $4$ dimensions, all invariant Lagrangians of a Weyl geometry with
curvature tensor $R_{bcd}^{a}$ and Weyl vector $W_{a}$ have been classified
\cite{Weyl}:
\[
c_{1}F_{ab}F^{ab}+c_{2}R^{2}+c_{3}R_{ab}R^{ab}+c_{4}R_{abcd}R^{abcd}
\]
The last term may be written as a linear combination of $R^{2}$, $%
R_{ab}R^{ab}$, and $C_{abcd}C^{abcd}$, where $C_{bcd}^{a}$ is Weyl's
conformal tensor defined by eq.(\ref{Conformal Tensor}), or it may be
eliminated altogether using the Gauss-Bonnet invariant. All of these actions
lead to higher order field equations. For example, Weyl's original free
action,
\[
S=\frac{1}{4}\int \sqrt{-g}(F_{ab}F^{ab}+R^{2})d^{4}x
\]
yields the fourth-order field equation \cite{Charap+Tait}
\[
R(R_{b}^{a}-\frac{1}{4}\delta _{b}^{a}R)+T_{b}^{a}=0
\]
As a result of this field equation, the metric is underdetermined. For
example, when $T_{b}^{a}=0,$ the single condition $R=0$ already provides a
solution. Since almost every metric is scale equivalent to one with $R=0,$
almost every metric is gauge equivalent to a solution.

Following the approach of MacDowell and Mansouri \cite{MacDowell+Mansouri}
for obtaining Einstein (super-)gravity through squaring the curvatures of
the de-Sitter group, Crispim-Romao, Ferber, and Freund \cite
{Romao+Ferber+Freund} and independently Kaku, Townsend, and van
Nieuwenhuizen \cite{Kaku+Townsend} derived Weyl (super-)gravity as a gauge
theory of the full conformal group. Gauging $O(4,2)$ under the conventional
constraints (\ref{Constraint 1}) and (\ref{Constraint 2}), vanishing torsion
and tracefree curvature, their $R_{abcd}R^{abcd}$-type Lagrangian reduces to
\[
C_{abcd}C^{abcd}=R_{ab}R^{ab}-\frac{1}{3}R^{2}
\]
All Weyl vector-dependent terms drop out of the action, whereas eq.(\ref
{Constraint 1}) renders the special conformal gauge field auxiliary. As in
the Chern-Simons case, it is given by $\mathcal{R}_{a}(\mathbf{e}^{a})$. All
other possible actions built out of the $O(4,2)$ curvatures under the
conventional constraints were shown to reduce to a Weyl geometry \cite
{Auxiliary Field}. It was concluded that Weyl's theory of gravity is the
unique conformally invariant gravity theory in $4$ dimensions.

These results were generalized to any dimension $n\geq 4$ by including a
compensating field \cite{Bergshoef}, so that the Lagrangian assumes the form
\[
e_{[a}^{\ \mu }e_{b}^{\ \nu }e_{c}^{\ \alpha }e_{d]}^{\ \beta }\phi ^{\frac{%
2(n-4)}{(n-2)}}R_{\mu \nu }^{\quad ab}R_{\alpha \beta }^{\quad cd}
\]
Under the conventional constraints (\ref{Constraint 1}) and (\ref{Constraint
2}) this reduces to
\[
(R_{ab}R^{ab}-\frac{n}{4(n-1)}R^{2})\phi ^{\frac{2(n-4)}{n-2}}
\]

As in the $4$-dim case, none of the quadratic action theories provide
obvious contact to Einstein gravity, but instead lead to higher derivative
field theories. Nonetheless, supersymmetrization of an $R^{2}$-action in $%
n=10$ \cite{Bergshoef} and $n=6$ \cite{BergSalam+Sezgin} is an important
issue which arises in the study of the low-energy limit of superstrings.

The comments of the preceeding two subsections regarding eq.(\ref{Curl of
conformal Ricci}) and the relationship between the special conformal gauge
field and $\mathcal{R}_{a}$ hold here as well (thought it should be noted
that eq.(\ref{Special conformal is curly R}) follows from the quadratic
field equations rather than only as a constraint). Thus, in contrast to
these quadratic-curvature theories, the linear biconformal theory:

\begin{itemize}
\item  provides direct contact with Einstein gravity,

\item  does not require compensating fields,

\item  does not require the conventional constraints (\ref{Constraint 1}) or
(\ref{Constraint 2})
\end{itemize}

Of course, biconformal space also permits curvature-quadratic actions for
any $n>2$ \textit{without} the use of compensating fields, although these
theories are not explored further here.

\section{Conclusion}

By finding the most general class of biconformal scale-invariant
curvature-poly-no\-mial actions for any dimension $n>2$, we have overcome
the well-known restrictions to the set of possible scale-invariant actions
in standard conformal gauge theory imposed by the coupling of the action to
the dimension, without the use of compensating fields. All of the displayed
polynomial actions rely on the existence of certain biconformally invariant
tensors as well as the scaling properties of the connection forms. Since the
solder and the co-solder forms that span the $2n$-dimensional biconformal
space scale with opposite weights, they provide a manifestly scale-invariant
volume element consisting of $n$\ solder and $n$\ co-solder forms. We also
displayed a Yang-Mills type scale invariant dual action, which hinges on the
existence of a scale-invariant biconformal dual operator.

For the most general linear action we computed and solved the field
equations by imposing them onto the minimal torsion biconformal structure
equations. With one exceptional case, all solutions have the following
properties:

\begin{enumerate}
\item  The full $2n$-dim space has a symplectic form, and is therefore
almost complex and almost K\"{a}hler.

\item  There are two $n$-dim involutions. The first leads to a foliation by
conformally flat manifolds spanned by weight $-1$ co-solder forms. The
second leads to a foliation by equivalent Ricci-flat Riemannian spacetimes
spanned by the weight $+1$ solder forms. The Riemann curvature is computed
from the solder form alone, despite the inclusion of minimal torsion,
general co-torsion and a general Weyl vector, and the spacetime is
Ricci-flat despite an arbitrary cosmological constant.

\item  The full $2n$-dim minimal torsion solutions are fully determined by
two fields, each defined entirely on the $n$-dimensional Riemannian
spacetimes: the solder form $e_{\mu }^{\ \ a}(x)\mathbf{d}x^{\mu }$ and a
symmetric tensor field, $k_{\alpha \beta }(x).$
\end{enumerate}

For the single special case, $\lambda =1,$ there still exists a foliation by
Ricci-flat Riemannian spacetimes, but the minimal field content includes the
one field beyond the solder form and $h_{a\alpha }$: the co-solder
coefficient $f_{a}^{\;\beta }$, which is coupled to $h_{a\alpha }$ by a
differential equation.

Certain important subclasses of biconformal spaces described in \cite{New
Conformal Gauging Paper} turned out to be special cases of the general
solution. Because of the symmetry between solder and co-solder form,
analogous results to the ones obtained hold for co-torsion-free biconformal
spaces, e.g. spacetime sector flatness. Spaces of vanishing torsion and
co-torsion are conformally flat.

\appendix

\section{Biconformal Gauging}

The conformal (M\"{o}bius) group $C(n)$\ is the group of transformations
preserving angles or ratios of infinitesimal lengths when acting on an $n$%
-dim space or, equivalently, leaving the null interval
\[
ds^{2}=\eta _{\mu \nu }dx^{\mu }dx^{\nu }=0
\]
with\allowbreak\ $\eta _{\mu \nu }=diag(-1,1\ldots 1)$, $\mu ,\nu =1\ldots n$%
\ , invariant. While $C(2)$\ is the infinite-dimensional diffeomorphism
group of the plane, the conformal group for $n>2$\ is a Lie group of
dimension $\frac{1}{2}(n+1)(n+2)$\ and locally isomorphic to the
pseudo-orthogonal group $O(n,2)$. It can be shown \cite{Barut} that $C(n)$\
is the projective group $O(n,2)/\{1,-1\}$\ of rays through the origin in $%
O(n,2)$. It possesses a real linear representation in $\mathbf{R}^{n+2}$, a
complex linear representation in $\mathbf{R}^{n}$, and a real nonlinear
representation in an $n$-dim compact spacetime (M\"{o}bius space).

Biconformal space was first introduced in \cite{New Conformal Gauging Paper}
using methods similar to the geometric construction of general relativity as
an ECSK theory. In the standard Poincar\'{e} gauge theory of gravitation one
postulates the invariance of some action integral under local Poincar\'{e}
transformations \cite{Utiyama/Kibble}. The field equations are derived by
``soldering'' the Lorentz fibers to the base manifold, i.e. by identifying
those gauge fields (connection forms) that correspond to the translation
generators of the Poincar\'{e} group with an orthonormal basis $\{\mathbf{e}%
^{a}\}$. This gauging was later recognized as being equivalent to Cartan's
orthonormal frame bundle formalism \cite{Cartan}. In this formalism, a
homogeneous space is first constructed as the quotient space of a group $G$\
and a subgroup with trivial core, i.e. a subgroup $G_{0}$\ that itself
contains no subgroup which is normal in $G$\ other than the identity. This
subgroup will act as the isotropy subgroup of any point in the orbit space $%
G/G_{0}$. The group action on this space is effective (only the identity of
the group acts as the identity transformation) and transitive (only one
orbit). The orbit space is a manifold with a Lie-algebra-valued connection
if the group is a Lie group. The affine connection of this frame bundle $%
\{\pi :G\rightarrow G/G_{0}\}$\ is then generalized to a Cartan connection
by including curvatures in the structure equations of the group. Holonomy
considerations require these curvatures to be horizontal, i.e. bilinear in
the base connections. The formalism provides \textit{a priori} locally group
symmetric geometries without requiring an action integral.

In this way, Minkowski space is built as the quotient space of the
Poincar\'{e} group acting on $\mathbf{R}^{4}$and the Lorentz group $O(3,1)$.
The curvature 2-forms associated with Lorentz transformations and
translations, Riemann curvature $\mathbf{R}_{b}^{a}$\ and torsion $\mathbf{T}%
^{a}$, respectively, are defined through the Poincar\'{e} structure
equations:
\begin{eqnarray*}
\mathbf{R}_{b}^{a} &=&\mathbf{d\omega }_{b}^{a}-\mathbf{\omega }%
_{b}^{c}\wedge \mathbf{\omega }_{c}^{a} \\
\mathbf{T}^{a} &=&\mathbf{de}^{a}-\mathbf{e}^{b}\wedge \mathbf{\omega }%
_{b}^{a}
\end{eqnarray*}
The resulting spacetime is a curved four-dimensional manifold with torsion.
The result can be generalized to $n$\ dimensions as well as applied to
manifolds with topology other than the usual $\mathbf{R}^{n}$\ topology.
\textit{Any} action constructed within this locally Poincar\'{e} invariant
geometry, such as the Einstein-Hilbert action
\[
S=\int \eta ^{a_{1}b}\mathbf{R}_{b}^{a_{2}}\wedge \mathbf{e}^{a_{3}}\wedge
\cdots \wedge \mathbf{e}^{a_{n}}\varepsilon _{a_{1}\ldots a_{n}}=\int \sqrt{%
-g}R\ d^{n}x
\]
provides a field theory.

In the frame bundle formalism language the standard conformal gauge theories
\cite{Romao+Ferber+Freund}-\cite{Bergshoef} correspond to a quotienting of
the conformal group by the inhomogeneous Weyl group generated by
Poincar\'{e} transformations and dilations \cite{Freund2}. While this
construction retains the largest possible continuous symmetry on the fibers,
it does not take the discrete symmetry of the conformal algebra into
account, according to which translation and co-translation generators are
essentially interchangeable.

Biconformal space is the $2n$-dimensional homogeneous space obtained by
quotienting the conformal group $C(n)$\ acting on M\"{o}bius space by the
homogeneous Weyl group $C_{0}$, consisting of Lorentz transformations and
dilations. Thus, symmetry of the fibers is exchanged for increased
coordinate freedom for the base manifold. In this gauging, the conformal
translation and co-translation generators are treated on an equal footing:
Their associated connection forms span the base space together. When broken
up into components based on their biconformal covariance properties, the $%
O(n,2)$ curvatures defined through the conformal structure equations (\ref
{Structure 1})-(\ref{Structure 4}) are bilinear in the these connection
forms. There are no \textit{a priori} conditions on the torsion and the
curvature.

\begin{definition}
\ Let $A,B,...=0,...,n$\ and $a,b,...=1,...,n$. A biconformal space is a
principal fiber bundle $\pi :C\rightarrow \mathcal{B}$\ with conformal
connection $\mathbf{\omega }_{B}^{A}=\{\mathbf{\omega }_{b}^{a},\mathbf{\
\omega }^{a},\mathbf{\omega }_{a},\mathbf{\omega }_{0}^{0}\}$, where $\pi $\
is the canonical projection of the $(n+1)(n+2)/2$-dimensional conformal
bundle onto the $2n$-dimensional base manifold $\mathcal{B}$\ induced by $%
C/C_{0}$, where the structure (or symmetry or gauge) group $C_{0}$\ is the
Weyl group of an $n$-dimensional Minkowski space.
\end{definition}

Biconformal space possesses a preferred orthonormal basis $\{\mathbf{\omega }%
^{a},\mathbf{\omega }_{a}\}$\ defined through the conformal Killing metric $%
g $\ , so that $g(\mathbf{\omega }^{a},\mathbf{\omega }^{b})=0$, $g(\mathbf{\
\omega }_{a}=0$, and $\mathbf{\omega }_{b})=0$\ and $g(\mathbf{\omega
}^{a},\mathbf{\
\omega }_{b})=\delta _{b}^{a}$. It provides a natural nondegenerate,
invariant $2$-form $\mathbf{\omega }^{a}\wedge \mathbf{\omega }_{a}$. In the
generic case of a biconformal space subject to the linear action and the
minimal torsion constraint discussed in this paper, the $2$-form is closed
and hence symplectic. By a well-known theorem \cite{Choquet-Bruhat} it is
always possible to construct on a symplectic manifold an almost complex
structure and a K\"{a}hler metric. Therefore, the field equations arising
from the linear action constrain the biconformal space to be almost
K\"{a}hler.

As a result of the increased dimension of biconformal space, there are many
new fields that could be identified with the electromagnetic potential or
other internal symmetries.

\section{The projected co-solder form, \textbf{f}$_{a}$}

We first define the projected co-solder form
\[
\mathbf{f}_{a}\equiv \mathbf{\omega }_{a|\mathbf{\omega }^{a}=0}
\]
Since the involution (\ref{Structure 2a}) allowed us to single out a set of $%
n$ biconformal coordinates $\{x^{\mu }\}$ for the weight $+1$ sector such
that $\mathbf{\omega }^{a}=e_{\mu }^{\ a}\mathbf{d}x^{\mu }$, we can find a
complimentary set of $n$ coordinates $\{z_{\mu }\}$ for the weight $-1$
sector so that $\mathbf{f}_{a}$ is of the form
\begin{equation}
\mathbf{f}_{a}=f_{a}^{\mu }(x,z)\mathbf{d}z_{\mu }  \label{f-1}
\end{equation}
This form shows that $\mathbf{f}_{a}$ is necessarily involute, since $%
\mathbf{df}_{a}$ is at least linear in $\mathbf{d}z_{\mu }$ and $\mathbf{d}
z_{\mu }=$ $f_{\mu }^{\;a}\mathbf{f}_{a}.$ On each $x_{0}=const.$
submanifold we can gauge $\omega _{b}^{a}$ and $\omega _{0}^{0}$ to zero
which implies $\mathbf{df}_{a}=0$ or, by the converse of the Poincar\'{e}
Lemma,
\[
\mathbf{f}_{a}=\mathbf{d}\theta _{a}
\]
for some 0-form $\theta _{a}(x_{0},z)$. When the last equation is extended
to the full biconformal space, $\theta _{a}$ becomes a function of $x$ and $%
z $, and therefore
\begin{eqnarray}
\mathbf{f}_{a} &=&\mathbf{d}\theta _{a}+\chi _{a\mu }\mathbf{d}x^{\mu }
\nonumber \\
&=&(\partial _{\mu }\theta _{a}+\chi _{a\mu })\mathbf{d}x^{\mu }+(\partial
^{\mu }\theta _{a})\mathbf{d}z_{\mu }  \label{f-3}
\end{eqnarray}
where $\partial _{\mu }$ and $\partial ^{\mu }$ denote derivatives with
respect to $x^{\mu }$ and $z_{\mu }$, respectively. Since by eq.(\ref{f-1}) $%
\mathbf{f}_{a}$ can have no part proportional to $\mathbf{d}x^{\mu },$ this
implies
\[
\chi _{a\mu }=-\partial _{\mu }\theta _{a}
\]
so that
\[
\mathbf{f}_{a}=(\partial ^{\mu }\theta _{a})\mathbf{d}z_{\mu }
\]
Defining a new set of coordinates by
\[
y_{\mu }\equiv e_{\mu }^{\;a}\theta _{a}
\]
and regarding $\theta _{a}$ as a function of $x$ and $y$, we have
\begin{eqnarray*}
\mathbf{f}_{a} &=&(\partial ^{\mu }\theta _{a})\mathbf{d}z_{\mu } \\
&=&(e_{a}^{\;\alpha }\frac{\partial y_{\alpha }}{\partial z_{\mu }})\mathbf{d%
}z_{\mu } \\
&=&e_{a}^{\;\beta }(\mathbf{d}y_{\beta }-\frac{\partial y_{\beta }}{\partial
x^{\alpha }}\mathbf{d}x^{\alpha })
\end{eqnarray*}
The partial derivative $\frac{\partial y_{\beta }}{\partial x^{\alpha }}$ is
computed holding $z_{\mu }$ constant. Writing this partial as a function $%
\psi _{\alpha \beta }(x^{\mu },y_{\nu })$, $\mathbf{f}_{a}$ takes the
desired form,
\[
\mathbf{f}_{a}=e_{a}^{\;\beta }\mathbf{d}y_{\beta }-\psi _{a\beta }\mathbf{d}%
x^{\alpha }
\]

\end{document}